\documentclass[useAMS,usenatbib]{mn2e}
\usepackage{lineno}
\def\aap{AA}
\def\aapr{AA Rev}
\def\apjl{ApJL}

\def\mnras{MNRAS}
\def\apj{ApJ}
\def\apjs{ApJS}
\def\aj{AJ}
\def\pasp{PASP}

\def\nat{Nat}

\def\NcontAA{15}
\def\NIncAA{10}

\def\jcap{JCAP}
\def\physrep{PhysRep}
\def\araa{ARA\&A}
\def\NlensTOT{7}
\def\NquadTOT{2}
\def\NNIQTOT{7}
\def\NIncTOT{27}
\def\NOBSTOT{117}
\newcommand{ \kms }{kms$^{-1}$}

\usepackage{bm}
\usepackage{graphicx}
\usepackage{float}
\usepackage{amssymb}
\usepackage{amsfonts}
\usepackage{amsmath}
\usepackage{color}
\usepackage{natbib}
\usepackage{hyperref}
\usepackage{longtable}
\usepackage{pdflscape}

\def\ucla{Department of Physics and Astronomy, PAB, 430 Portola Plaza, Box 951547, Los Angeles, CA 90095-1547, USA}
\def\eso{European Southern Observatory, Karl-Schwarzschild-Strasse 2, 85748 Garching bei M{\"u}nchen, DE}
\def\kipac{Kavli Institute for Particle Astrophysics and Cosmology, Stanford University, 452 Lomita Mall, Stanford, CA 94305, USA}
\def\valpo{Instituto de Física y Astronomía, Universidad de Valparaíso, Avda. Gran Breta{\~n}a 1111, Playa Ancha, Valparaíso 2360102, Chile }
\def\abello{Departamento de Ciencias Fisicas, Universidad Andres Bello Fernandez Concha 700, Las Condes, Santiago, Chile}

\def\asiaa{Institute of Astronomy and Astrophysics, Academia Sinica, P.O. Box 23-141, Taipei 10617, Taiwan}
\def\mpa{Max-Planck-Institut f\"ur Astrophysik, Karl-Schwarzschild-Str. 1, D-85741 Garching, Germany}
\def\ucd{Department of Physics, University of California Davis, 1 Shields Avenue, Davis, CA 95616, USA}
\def\asiaa{Institute of Astronomy and Astrophysics, Academia Sinica, P.O.~Box 23-141, Taipei 10617, Taiwan}
\def\ioa{Institute of Astronomy, Madingley Road, Cambridge CB3 0HA, UK}

\def\mit{MIT Kavli Institute for Astrophysics and Space Research, 37-664G, 77 Massachusetts Avenue, Cambridge, MA 02139}

\def\epfl{Laboratoire d'Astrophysique, Ecole Polytechnique F\'ed\'erale de Lausanne (EPFL), Observatoire de Sauverny, CH-1290 Versoix, Switzerland}

\def\ports{Institute of Cosmology \& Gravitation, University of Portsmouth, Portsmouth, PO1 3FX, UK}

\def\ufrgs{Departamento de Astronomia, Instituto de Fsica da Universidade Federal do Rio Grande do Sul, 91501-970, Porto Alegre, Brazil}

\def\milchile{Millenium Institute of Astrophysics, Chile}
\def\tum{Physik-Department, Technische Universit\"at M\"unchen, James-Franck-Stra\ss{}e~1, 85748 Garching, Germany}
\def\subaru{Subaru Telescope, National Astronomical Observatory of Japan, 650 North A'ohoku Place, Hilo, HI 96720, U.S.A.}
\def\ttemail{\tt tt@astro.ucla.edu}

\title[STRIDES-I]{The STRong lensing Insights into the Dark Energy Survey (STRIDES) 2016 follow-up campaign. I. Overview and classification of candidates selected by two techniques.}
\author[Treu et al.]{\parbox{\textwidth}{
  T.~Treu$^{1}$\thanks{\ttemail},
  A.~Agnello$^{1,2}$,
  M.A.~Baumer$^{3}$,
  S.~Birrer$^{1}$,
  E.J.~Buckley-Geer$^{4}$,
  F.~Courbin$^5$,
  Y.J.~Kim$^{3}$,
  H.~Lin$^{4}$,
  P.J.~Marshall$^{3}$,
  B.~Nord$^{4}$,
  P.L.~Schechter$^6$,
  P.R.~Sivakumar$^{4,7,8}$,
L.E.~Abramson$^{1},$
 T.~Anguita$^{9,17}$,
 Y.~Apostolovski$^9$,
M.W.~Auger$^{10}$,
 J.H.H.~Chan$^{7,11}$,
 G.C.F.~Chen$^{12}$,
T.~E.~Collett$^{13}$,
 C.D.~Fassnacht$^{12}$,
 J.-W.~Hsueh$^{12}$,
C.~Lemon$^{10}$,
R.G.~McMahon$^{10}$,
 V.~Motta$^{14}$,
F.~Ostrovski$^{10,15}$,
 K.~Rojas$^{14}$,
 C.E.~Rusu$^{12,19}$,
 P.~Williams$^{1}$,
J.~Frieman$^{4}$,
G.~Meylan$^{5}$,
S.H.~Suyu$^{16,11,18}$,
T.~M.~C.~Abbott$^{20}$,
F.~B.~Abdalla$^{21,22}$,
S.~Allam$^{4}$,
J.~Annis$^{4}$,
S.~Avila$^{13}$,
M.~Banerji$^{10,23}$,
D.~Brooks$^{21}$,
A.~Carnero~Rosell$^{24,25}$,
M.~Carrasco~Kind$^{26,27}$,
J.~Carretero$^{28}$,
F.~J.~Castander$^{29,30}$,
C.~B.~D'Andrea$^{31}$,
L.~N.~da Costa$^{24,25}$,
J.~De~Vicente$^{32}$,
P.~Doel$^{21}$,
T.~F.~Eifler$^{33,34}$,
B.~Flaugher$^{4}$,
P.~Fosalba$^{29,30}$,
J.~Garc\'ia-Bellido$^{35}$,
D.~A.~Goldstein$^{36,37}$,
D.~Gruen$^{3,38}$,
R.~A.~Gruendl$^{26,27}$,
G.~Gutierrez$^{4}$,
W.~G.~Hartley$^{39,21}$,
D.~Hollowood$^{40}$,
K.~Honscheid$^{41,42}$,
D.~J.~James$^{43}$,
K.~Kuehn$^{44}$,
N.~Kuropatkin$^{4}$,
M.~Lima$^{45,24}$,
M.~A.~G.~Maia$^{24,25}$,
P.~Martini$^{41,46}$,
F.~Menanteau$^{26,27}$,
R.~Miquel$^{28,47}$,
A.~A.~Plazas$^{34}$,
A.~K.~Romer$^{48}$,
E.~Sanchez$^{32}$,
V.~Scarpine$^{4}$,
R.~Schindler$^{38}$,
M.~Schubnell$^{49}$,
I.~Sevilla-Noarbe$^{32}$,
M.~Smith$^{50}$,
R.~C.~Smith$^{20}$,
M.~Soares-Santos$^{51}$,
F.~Sobreira$^{52,24}$,
E.~Suchyta$^{53}$,
M.~E.~C.~Swanson$^{27}$,
G.~Tarle$^{49}$,
D.~Thomas$^{13}$,
D.~L.~Tucker$^{4}$,
A.~R.~Walker$^{20}$}}
\begin{document}
\voffset-.6in

\date{Accepted . Received }

\pagerange{\pageref{firstpage}--\pageref{lastpage}}

\maketitle

\label{firstpage}

\begin{abstract}
  The primary goals of the STRong lensing Insights into the Dark
  Energy Survey (STRIDES) collaboration are to measure the dark energy
  equation of state parameter and the free streaming length of dark
  matter. To this aim, STRIDES is discovering strongly lensed quasars
  in the imaging data of the Dark Energy Survey and following them up
  to measure time delays, high resolution imaging, and spectroscopy
  sufficient to construct accurate lens models. In this paper, we
  first present forecasts for STRIDES. Then, we describe the STRIDES
  classification scheme, and give an overview of the Fall 2016
  follow-up campaign.  We continue by detailing the results of two
  selection methods, the Outlier Selection Technique and a
  morphological algorithm, and presenting lens models of a
  system which could possibly be a lensed quasar in an unusual
  configuration. We conclude with the summary statistics of the Fall
  2016 campaign. Including searches presented in companion papers
  (Anguita et al.; Ostrovski et al.), STRIDES followed up \NOBSTOT\
  targets identifying \NlensTOT\ new strongly lensed systems, and
  \NNIQTOT\ nearly identical quasars (NIQs), which could be confirmed as
  lenses by the detection of the lens galaxy.  76 candidates were
  rejected and \NIncTOT\ remain otherwise inconclusive, for a success
  rate in the range 6-35\%. This rate is comparable to that of
  previous searches like SQLS even though the parent dataset of
  STRIDES is purely photometric and our selection of candidates cannot
  rely on spectroscopic information.
\end{abstract}

\section{Introduction}
\label{sect:intro}


In the four decades since the discovery of the first strongly lensed
quasars \citep{WCW79,Wey80}, they have morphed from an intellectual
curiosity to a powerful and in some sense unique astrophysical tool
\citep{CSS02}. Three classes of applications make strongly lensed
quasars especially valuable. First, by modeling how the light of the
background quasar and its host galaxy is distorted one can reconstruct
the distribution of luminous and dark matter in the deflector, and
thus address fundamental astrophysical problems like the normalization
of the stellar initial mass function \citep{Poo++09,Sch++14} and the
abundance of dark matter subhalos
\citep{M+S98,M+M01,D+K02,Nie++14,Nie++17,BAR17}. Second, by exploiting
magnification, one can study in great detail the distant quasars, the
properties of their accretion disks and host galaxies
\citep{Peng:2006p236,Din++17}. Third, by measuring time delays between
the variable images and stellar kinematics of the deflector one can
measure cosmic distances and thus cosmological parameters, especially
the Hubble Constant
\citep{Ref64,Sch++97,T+K02b,S+S13,Suy++13,T+M16,Suy++14,BAR16,Bon++17,T+K18,STA18}.

Unfortunately, most applications to date have been limited by the
small number of known suitable systems. Lensed quasars, patricularly
the ones with four images which provide the most information, are rare
on the sky \citep[of order 0.1 per square degree at present day
typical survey depth and resolution;][]{O+M10}. Therefore successful
searches for lensed quasars require searches over large solid angles
\citep[e.g.,][]{Bro++03,Ina++12,Mor++16} and substantial follow-up to
weed out false positives.

Furthermore, not every lensed quasar system is suitable for every
application: depending on the specifics of the lensing configuration
and on the brightness of deflector and source, some systems contain
significantly more information than others. Thus, in practice, every
application of strongly lensed quasars to date has been limited to
samples of one or two dozens at best.

The present generation of wide field imaging surveys provides an
opportunity to make transformative measurements by increasing
dramatically the sample of known lens quasars. Hundreds of strongly
lensed quasars are hiding in the thousands of square degrees currently
being imaged by the Dark Energy Survey (and similarly, e.g., the
Hyper-Suprime-Cam SSP Survey, the VST-ATLAS Survey), waiting to be
discovered and followed up.

In order to exploit the bounty of data provided by DES, we have formed
the STRIDES collaboration (STRong lensing Insights into Dark Energy
Survey \footnote{STRIDES is a Dark Energy Survey Broad External
  Collaboration; PI: Treu. \url{http://strides.astro.ucla.edu}}). The
immediate goal of STRIDES is to identify and follow-up large numbers
of multiply imaged quasar motivated by two main science drivers: i)
study dark energy using gravitational time delays; ii) study dark
matter using flux ratio and astrometric anomalies. Additional science
goals include the determination of the normalization of stellar
mass-to-light ratios of massive early-type galaxies and the
properties of accretion disks through the study of quasar
microlensing \citep[e.g.,][]{Mot++17}.

As we will show in this paper, STRIDES can in principle discover
enough strongly lensed quasars to make substantial headway on both its
two main science drivers.  Strongly lensed quasars' main contribution
to dark energy measurements is through the determination of absolute
distances in the relatively low redshift universe, and thus of the
Hubble Constant $H_0$ \citep{T+M16}. In turn, knowledge of $H_0$ is
currently a limiting factor in the interpretation of cosmic microwave
background data \citep{Wei++13,Bon++17}. Current measurements of $H_0$
based on the local distance ladder method reach $\sim 2.4$\% precision
\citep{Rie++16,Rie++18a,Rie++18b}. The most recent time delay based measurements reach
$\sim3.8$\% with just 3 systems \citep{Bon++17}.  Reaching 1\%
equivalent precision on H$_0$ is extremely important
\citep{Suy++12,Wei++13,T+M16} and it will require $\sim40$ lensed
quasars \citep{JeeEtal2016,STA18} with data and models of quality
equivalent to those presented by
\citet{Suy++17,Rus++17,Slu++17,Won++17,Bon++17}. Similarly, current
limits on dark matter substructure are based on $\sim10$ lenses
\citep{D+K02,Veg++14,Nie++17}. Quadrupling the sample of viable quads
will be a major step forward in constraining the properties of dark
matter \citep{Gil++18}.

Finding lensed quasars in purely imaging datasets of the size of DES
is an unprecedented task. It requires the development of new
algorithms to identify candidates from the imaging data, and
substantial investment of telescope time to follow-up and confirm the
candidates.  In order to maximize completeness and purity, the
collaboration is pursuing multiple independent approaches to identify
candidate lenses.  The lack of $u$-band imaging data in DES makes it
particularly hard to identify QSOs, therefore many of the selection
techniques combine DES imaging with WISE photometry.  The candidates
are then followed up with spectroscopy and higher resolution
imaging. Both are necessary to confirm the lensing nature of the
systems and obtain the redshift and astrometry necessary for modeling
and scientific exploitation. First results from the STRIDES program
have been presented by \citet{Agn++15,Lin++17}. Once the candidates
are confirmed, the best ones are selected for monitoring either with
the 1.2m Euler Telescope or with the MPIA 2.2m Telescope at La Silla
as part of the COSMOGRAIL network \citep{Cou++18}.

This paper has multiple aims. First, it provides an overview of the
STRIDES program and forecasts the number of expected lensed quasars to
be found in the complete Dark Energy Survey (DES;
\S~\ref{sec:forecasts}). The forecasts show that the DES area depth
and resolution should be sufficient to more than double the current
sample of known lensed quasars, providing new systems especially in
the South hemisphere outside the area covered by previously largest
search based on the Sloan Digital Sky Survey \citep{Ina++12}. Second,
this paper defines a candidate classification system, and various
subclasses of inconclusive and contaminant sources
(\S~\ref{sec:class}). The system will be applied throughout the
collaboration with the goal to standardize the lens confirmation
process and hopefully adopted by other investigators. Third, this
paper gives an overview of the Fall 2016 Follow-up campaign
(\S~\ref{sec:overview}), listing the candidates selected by two
techniques (\S~\ref{sec:outliers} and \S~\ref{sec:morphological}) that
did not yield any confirmed lens, except for a possible unusual
quadruply imaged quasar. Companion papers in this series present the
follow-up of candidate lensed QSOs selected using other techniques
\citep[][and Ostrovski et al. 2018, in preparation]{Ang++18}, showing
spectra and images for all confirmed lenses and otherwise promising
inconclusive systems\footnote{During the follow-up campaign, non-DES
  targets selected from other surveys were also targeted. Those are
  described by papers outside of this series
  \citep[e.g.,][]{Sch++17,Wil++18,Ost++18,Agn++18b}.}. The fourth goal
of this paper is to present the summary statistics of the 2016
follow-up campaign, combining the results from every search method, as
discussed in \S~\ref{sec:ss}. Target selection for the 2016 campaign
was based on early DES datasets, which did not cover the full depth
and footprint of the survey. Thus, the Fall 2016 campaign statistics
are not sufficient for a detailed comparison with the forecast for
STRIDES. However, the follow-up statistics are sufficient for an
assessment of the success rate and the completeness of the searches so
far. Remarkably, the success rate is comparable to those of previous
searches, even though no spectroscopic preselection or u-band imaging
was available. A short summary concludes the paper in
\S~\ref{sec:summary}.

All magnitudes are given in the AB system, and a standard concordance
cosmology with $\Omega_m=0.3$, $\Omega_\Lambda=0.7$, and $h=0.7$ is
assumed when necessary.

\section{Forecasts for STRIDES}
\label{sec:forecasts}

Our forecasts for STRIDES use the OM10 mock lensed quasar catalog of
\citet{O+M10}. The reader is referred to the original paper and
associated software
repository\footnote{\texttt{https://github.com/drphilmarshall/OM10/}}
for full details of how this basic catalog was generated. Here we give
only a concise summary for the convenience of the reader. The
deflector population is assumed to consist of early-type galaxies,
which represent 80-90\% of the galaxy-scale lenses
\citep{TOG84,Bol++08a} and dominate the optical depth for image
separations in the range $0\farcs5-4''$ considered here. We do not
consider systems with smaller image separation since they would be
unresolved in the DES images. Systems with image separation larger
than $4''$ would be dominated by group- and cluster-scale lenses, and
thus are not appropriately captured by the OM10 framework.

The deflector population is described by the velocity dispersion
function of early-type galaxies \citep{CPV07}, which has been shown to
reproduce well the abundance of strong lenses
\citep{Cha07,Oguri:2008p277}. The deflector potential is described by
a single isothermal ellipsoid \citep{Kormann:1994p186}, which is the
simplest model that gives a sufficiently accurate description of
early-type galaxies \citep{Tre10}, with external shear to account for
the contribution of the environment along the line of sight
\citep{KKS97}. Multiband fluxes based on the observed correlation
between the velocity dispersion and luminosity of early-type galaxies
\citep{HydeBernardi2009} are computed using the publicly available
code {\sc
  LensPop},\footnote{\texttt{https://github.com/tcollett/LensPop}}
written by one of us \citep[T.C.][]{Collett2015}.

The quasar source population is described by a redshift-dependent
double power law luminosity function consistent with SDSS data
\citep{Fan++01,Richards:2006p1587,Ric++05}.

\begin{figure}
  \centering \includegraphics[angle=0, width=0.95\linewidth]{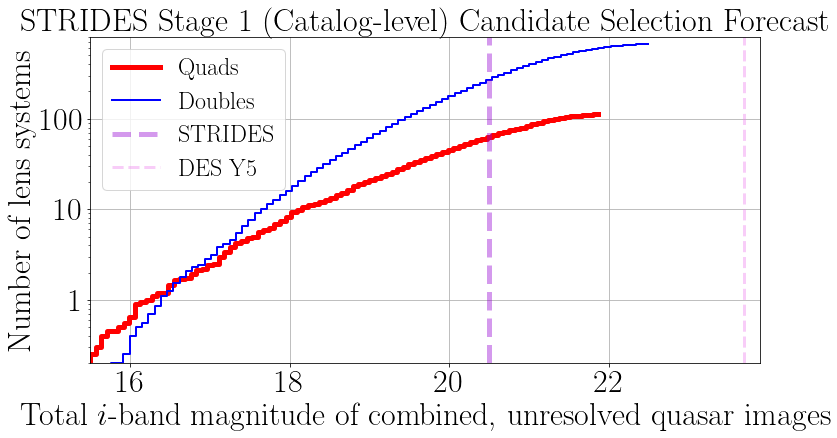}
  \centering \includegraphics[angle=0, width=0.95\linewidth]{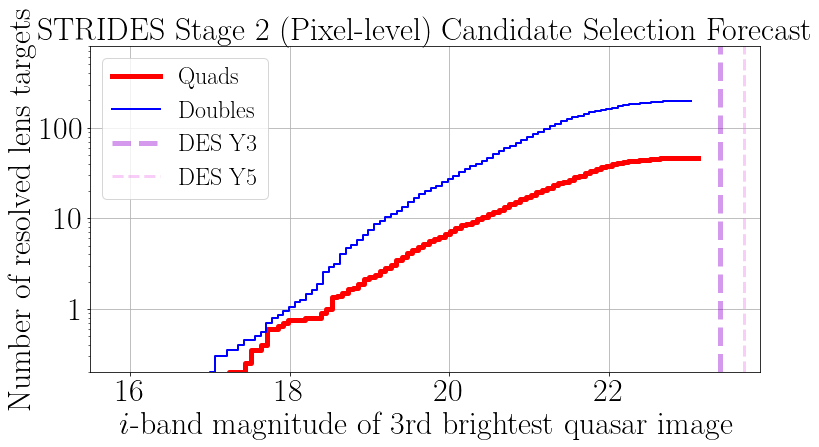}
  \caption{Cumulative number of expected lensed quasars for STRIDES based on the
    \citet{O+M10} catalog, in an approximate emulation of the STRIDES lens selection. Thick solid red lines represent quads, thin solid blue lines represent doubles.  Top: expected number of lenses in the DES+WISE catalogs, as a function of total lensed quasar magnitude. Bottom: expected number of catalog-selected lenses visible (i.e. resolved as lenses, with detectable counter-images) in the DES images.}
\label{fig:forecasts1}
\end{figure}

Figure~\ref{fig:forecasts1} shows the expected number of multiply-imaged
quasars as a function of total (unresolved) quasar $i$-band
magnitude, given the 5000 square degree footprint. By combining all
the lensed quasar light to compute each mock lens' total quasar
brightness, we enable an approximate emulation of a catalog-level
selection in which no lens system is resolved into component multiple
images, as is certainly the case for the WISE photometry (primarily W1 and W2,
with resolution $\sim6''$) that we use in our candidate
selection. The number counts curves are truncated because the OM10
catalog was generated so as to contain lens systems whose 3rd
brightest image would be detected at 10-sigma in a single LSST visit,
with depth $i=24.5$. This gives a mock sample bright enough for our
purposes.

From Figure~\ref{fig:forecasts1} we see that quad systems outnumber
double systems at magnitudes brighter than $i\approx 16.6$. While our
actual photometric joint DES+WISE catalog selection is complex, it
leads to a ``Stage 1'' list of candidates that is
significantly incomplete below a total quasar magnitude of
$i \approx 20.5$. Above this limit, we expect the DES survey area to
contain about 60 quads and 250 doubles.

However, many of these systems will have multiple image separations
that are too small to be resolved, and counter-images that are too
faint to see, and so we expect these to be ranked lowly in any imaging
follow-up campaign. Requiring that the image separation be greater
then $0\farcs9$ to be resolved in the DES survey images, and that the
2nd (in doubles) or 3rd (in quads) brightest image to be detected
above the DES Y3 detection limit of $i = 23.4$ emulates a ``Stage 2''
image inspection selection, that leads to a reduced number of {\it
  visibly} multiply-imaged lensed quasars. We see from the lower panel
of Figure~\ref{fig:forecasts1} that we expect this Stage 2 sample to
contain about 50 quads and 200 doubles, for a total of about 250
potentially detectable lens systems.

To put the STRIDES forecasts in context, this Stage 2 sample is larger
than all the currently published lensed quasars, i.e. approximately 40
quads and 140 doubles, including cluster scale
deflectors.\footnote{Compilation assembled by one of us (C.L.). The
  compilation by \citet{Duc++18}, not yet publicly available, reports
  243 confirmed systems, even though a direct comparison is difficult
  since the criteria for confirmation are unpulished and may be
  different from ours. Regardless of which compilation one chooses to
  compare, the STRIDES forecasted sample is larger than the number of
  currently known lensed quasars} We can also compare the STRIDES
forecasts to the outcome of the Sloan Digital Sky Survey (SDSS)
searches and the expectations and results for the recently completed
Kilo Degree Survey \citep[KIDS;][]{deJ++03}.  The Sloan Digital Sky
Survey Quasar Lens Search \citep[SQLS;][]{Ina++12} reported 26 lensed
quasars as part of their statistical sample (6 were known prior to
SQLS), plus an additional 36 found with a variety of techniques (14
were known prior to SQLS). Of the 26 quads in the statistical sample,
4 are galaxy-scale quads (including one previously known) and one is a
cluster-scale 5 image lens.  Of the non-statistical sample, 5 systems
have 4 or more images (4 of which were previously known). The
statistical sample is limited to a total quasar brightness of
$i<19.1$, and the non-statistical sample extends to $i\sim20$. As
described above, DES should be able to deliver a significantly larger
number of lenses by virtue of the superior depth and resolution of its
images, even though the area covered on the sky is smaller than that
of SDSS. Furthermore, the overlap between SDSS and DES is minimal, so
the two searches are complementary in terms of sky coverage and
follow-up opportunities.  Similarly to DES, KIDS targets the southern
hemisphere, but its smaller solid angle coverage limits the yield in
terms of lenses. An approximate forecast can be obtained by scaling
the DES predictions by the ratio of the sky coverage. Thus, KIDS Data
Release 3 and 4 should find approximately 10\% and 20\% of the lenses
present in DES \citep{Spi++18}.

\begin{figure}
  \centering
  \includegraphics[angle=0, width=0.5\textwidth]{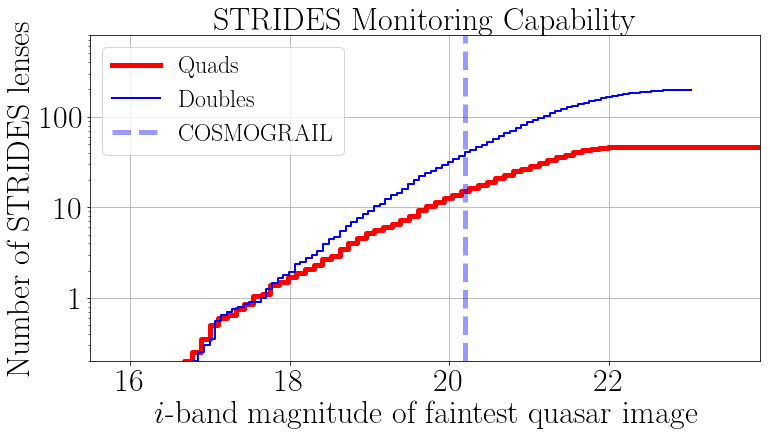}
  \caption{Cumulative number of expected lensed quasars for STRIDES based on
    \citet{O+M10} catalog as a function of their faintest quasar image's magnitude. Thick solid red lines represent quads, thin solid blue lines represent doubles. The practical limit for current monitoring capabilities are shown as a vertical dashed line. }
\label{fig:forecasts2}
\end{figure}

An important caveat for the use of the sample for time delay
cosmography is that the brightness of the faintest image in the system
is the limiting factor for monitoring.  In Figure~\ref{fig:forecasts2}
we show the predicted number of lenses as a function of faintest image
magnitude. From our COSMOGRAIL lens monitoring experience, we expect
to be able to measure the time delay well for systems with faintest
image brighter than $i\approx 20.2$ and image separation larger than
$0\farcs9$, with the current allocation of 1m/2m class telescopes
\citep[going fainter would require more time on 1m/2m telescopes or
upgrading to a 4m telescope;][]{Tre++13}. This practical limit leads
to a prediction of there being about 15 quads and 45 doubles bright
enough to monitor well in the DES area, with current monitoring
capabilities.  Exploiting the statistical power of the larger STRIDES
sample of 50 quads and 200 doubles will require monitoring on 4m class
telescopes, or much larger time allocations on the 1-2m class
telescopes that are currently used. For example, COSMOGRAIL is
currently using 20\% of the time on a 2.2m telescope \citep{Cou++18}.


\begin{figure}
  \centering
  \includegraphics[angle=0, width=0.5\textwidth]{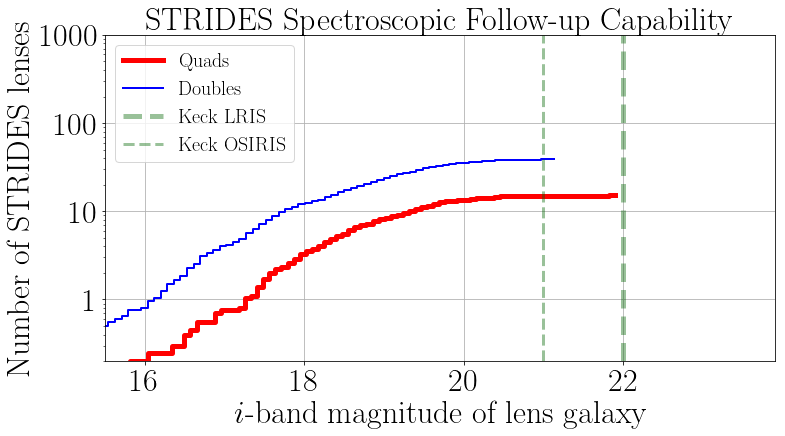}
  \caption{Cumulative number of expected lensed quasars for STRIDES based on
    \citet{O+M10} catalog as a function of deflector magnitude. Thick solid red lines represent quads, thin solid blue lines represent doubles. The practical limit for velocity dispersion measurement based on current instruments on 8-10m ground based telescopes are shown as vertical dashed lines.}
\label{fig:forecasts3}
\end{figure}

The deflector magnitude is also an important consideration for
constructing a cosmographic time delay lens sample, since it is a
limiting factor in the determination of stellar kinematics used to
break the mass-sheet degeneracy. The forecast as function of deflector
magnitude is shown in Figure~\ref{fig:forecasts3}.
We compute lens galaxy magnitudes using the stellar population
synthesis code provided with the {\sc LensPop}
package~\citep{Collett2015}. For a given velocity dispersion, we
compute the absolute rest-frame $r$-band magnitude using the relation
of \citet{HydeBernardi2009}. This is then converted to observed
apparent magnitudes using the redshift of the lens, a flat $\Lambda$CDM
cosmology with $h=0.7$ and $\Omega_{\rm M}=0.3$, and assuming a 9~Gyr
old population with solar metallicity.
We use the LRIS and OSIRIS spectrographs at the Keck Observatory for
the stellar kinematics measurements: we see that with these facilities
we should expect {\it all} of our STRIDES systems to have easily
measured lens galaxy stellar kinematics to 6-7\% precision, which is
the current state of the art in this field \citep{Won++17}.

\begin{figure*}
  \centering
  \includegraphics[angle=0, width=\textwidth]{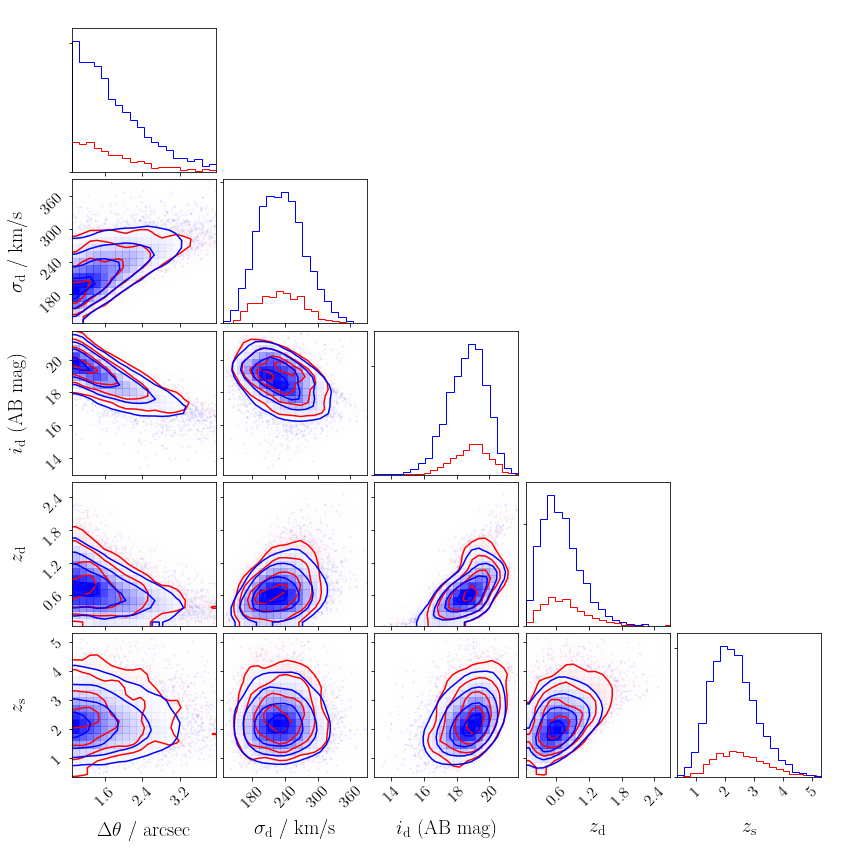}
  \caption{Distribution of basic properties for the expected lensed
    quasar sample for STRIDES based on the \citet{O+M10} catalog. As in the other panels in this section, the more numerous doubles are shown in blue, while quads are shown in red. The histograms on the diagonal show one dimensional normalized probablity distribution functions based on the samples.}
\label{fig:forecasts4}
\end{figure*}

Other interesting properties of the expected sample are shown in
Figure~\ref{fig:forecasts4}. We expect the deflector redshift
distribution to peak at around $z_d\sim0.5$, while the sources peak at
redshifts between $z_s\sim2-3$. As expected, the distribution of
velocity dispersion of the deflector peaks around $\sigma \sim$250
\kms, due to the exponential cutoff of the velocity dispersion
function for large $\sigma$ and the $\sigma^4$ dependency of lensing
cross section.

\section{STRIDES classification scheme and criteria}
\label{sec:class}

In this section we define the STRIDES classification scheme for
confirmed lenses, inconclusive systems, and contaminants. Although
some degree of subjectivity is inherent in this classification, we
will strive to keep a consistent classification scheme throughout the
STRIDES follow-up campaigns in 2016 and future years. In general, we
find that both high resolution imaging and spectroscopy are necessary
to classify a system as a definite lens, except in the following
cases: imaging is sufficient if arcs are detected or if the
configuration is consistent with a classic quad configuration (cross,
fold, cusp). Spectroscopy is sufficient if the quasar spectra are
partially resolved and a deflector galaxy is detected in between.

\subsection{Secure, probable, and possible lenses}

We adopt a different classification scheme for doubles and quads, as
detailed below.

\subsubsection{Confirmed lenses}

For doubles, we require for confirmation one of the following
scenarios: 1) nearly identical spectra of the two quasar images
(i.e. differences consistent with microlensing and differential dust
extinction), as well as the detection of the lens galaxy in imaging or
spectroscopy; 2) the detection of gravitationally lensed arcs
consistent with extended images of the quasar host galaxy; 3) coherent
and delayed variability (although in general the DES data do not
sample enough epochs to classify according to this criterion).

For quads, given the intrinsic rarity of the configuration (and
possible contaminants) and the difficulty of detecting the deflector
galaxy and separating the lensed quasar images in spectroscopy, we
only require spectroscopic confirmation of at least one of the quasar
images, the detection of flux consistent with a lens galaxy, plus a
configuration consistent with strong lensing. The latter could
consist, e.g. of three point sources, plus an extended source
interpreted as the fourth image blended with the deflector, or other
similar configurations.

\subsubsection{Probable and possible quads}

We use these categories for candidate quads that do not fullfill all
criteria for confirmation. Again, this classification is inherently
subjective, involving some personal judgement of what is a satisfactory
lens model and lens galaxy detection. It is similar in spirit to the
classification (secure, probable, possible) adopted by the SLACS Team
\citep{Bol++08, Aug++09}. In an attempt to quantify the degree of
certainty we consider secure systems that have 99\% probability of
being lenses, probable at 95\%, and possible at 68\%, whenever such
quantification is possible.

\subsection{Inconclusive systems}

Inconclusive systems include candidate doubles that have a significant
likelihood of being lenses based on the available good quality data,
and those for which the data are just insufficient to make any
statement.

\subsubsection{Nearly Identical Pairs of Quasars (NIQs)}

We call nearly identical quasars (NIQ), every pair of
spectroscopically confirmed quasars, for which spectral differences
can be explained via microlensing or differential extinction, but
there is no detection of a deflector galaxy. The non-detection can be
due to insufficiently deep high resolution imaging data or
spectroscopy, or could be due to the system being composed of an
actual physical pair as opposed to two lensed images.  We single out
this class of systems as primary targets for additional
follow-up. Conversely, if the data are deep enough to rule out the
presence of a lens galaxy, estimated in the following manner, the
system is classified as not a lens. First, we take half the image
separation as best estimate of the Einstein Radius. Second, as a
function of source and deflector redshift, the Einstein Radius is
transformed into stellar velocity dispersion $\sigma_*$, adopting a
singular isothermal sphere model with normalization $\sigma_{\rm SIS}$
equal to $\sigma_*$ \citep[e.g.,][]{Tre++06}. We limit the range of
acceptable redshifts to those that yield $\sigma_*<500$ \kms. Third,
we assign an apparent magnitude to each set of deflector redshift
$\sigma_*$ using the empirical relation for early type-galaxies given
by \citet{Mas++15}. Fourth, we adopt the maximum of the possible
magnitudes as faintest possible flux from the source. This magnitude
is typically fainter than what we can reach in our standard ground
based follow-up and usually requires deep/high resolution data from
HST or adaptive optics, since the lens galaxy could be hiding under
the bright quasar in ground based seeing limited images.

\subsubsection{Otherwise Inconclusive Doubles}

This class contains all other cases of candidate doubles where the
quality of the data is insufficient to confirm or rule out the lensing
hypothesis. This may include systems that look like doubles but there
is spectroscopic confirmation of just one putative image (or blended),
or systems that are consistent with doubles in higher resolution
imaging (from space or adaptive optics), but do not show arcs or have
sufficiently good spectroscopy to confirm them as lenses.

\subsection{Contaminants}

Whenever possible we classify false positive doubly imaged quasars in
one of the following, mostly self-explanatory, classes: 1) Quasar-Star
Pair; 2) Quasar Pair; 3) Quasar-Galaxy Pair, either based on
spectroscopic classification or based on one of the two candidate
images being resolved at resolution higher than that of the discovery
images and inconsistent with lensed arcs; 4) Galaxy Pair or
Merging/Irregular Galaxy, either based on spectroscopic classification
or based on the two candidate images being resolved at resolution
higher than that of the discovery image and inconsistent with lensed
arcs; 5) Other.

\section{Overview of the Fall 2016 Follow-up campaign}
\label{sec:overview}

Based on the definitions introduced in the previous section, for
definite confirmation we require the spectra of the multiple imaged
quasars to be almost identical, although not exactly the same in order
to allow for differences due to variability, microlensing, and line of
sight effects. In addition, we require the detection of a main
deflector galaxy, either photometrically or through foreground
spectroscopic features.  Alternatively, high resolution imaging alone
is sufficient if the system is a classic quad or arcs are detected.

Motivated by the goal to identify/reject as many candidates as
possible we applied for telescope time for both adaptive optics
imaging and spectroscopy. We applied to 3m-10m class telescopes chosen
based on instrument configuration and time availability during
semester 2016B, such that the entire DES footprint was available. Our
proposals were successful even though we had limited control over when
the runs were scheduled. The criteria of the target choice and
scheduling for each run are given below.

The telescopes and instruments used during the campaign and the dates
of each run are summarized in Table~\ref{tab:or}.  The 3.6m New
Technology Telescope at La Silla was used primarily for spectroscopy.
EFOSC2 was used with the \#13 grism and $1\farcs2$ wide slit. The
detector was binned two by two, resulting in a dispersion of
5.44\AA/pixel, a pixel scale of $0\farcs24$ along the slit, and
wavelegth coverage from 3685 to 9315\AA. Typically one or two
exposures of 600s were taken for each object.  The 4.1m Southern
Astrophysical Research Telescope at Cerro Pachon was used primarily
for high resolution imaging with its Adaptive Optics (AO) system
SAM. Imaging was taken with the CCD SAMI through the $z$-band to
maximize AO correction and optmize the contrast between quasar and
deflector galaxy. The pixel scale was 0$\farcs$045/pixel (with a
2$\times$2 binning yielding 0$\farcs$09/pixel) and the typical
exposure time was 3x180 seconds.  When the weather was not conducive
to AO imaging, SOAR was used with the Goodman Spectrograph to take
spectra.  For the Goodman setup we used the 400 lines/mm grating with
the blocking filter GG455, with a binning of 2$\times$2 and a slit
width of 1 arcsec.  Between 2 and 3 exposures of 1200 seconds were
taken per target.

The 10m Keck-2 Telescope was used to follow-up the candidates visible
from Maunakea, both for spectroscopy with the Echellette Spectrograph
and Imager (ESI) and for imaging with the Near InfraRed Camera 2
behind the adaptive optics system.  ESI was used in the default
Echellette mode with the $1''$ arcsecond slit, while NIRC2 was used in
the narrow field configuration (10mas pixel) in the K-band in order to
maximize AO correction. One adaptive optics run was scheduled on the
3m Shane Telescope aimed for the brighter candidates but was lost to
weather.

Outside of the main campaign, a few images and spectra were obtained
with the 6.5 Magellan Telescopes. Those will be discussed in the
appropriate context in papers II and III. $i$-band images of candidate
DESJ2346-5203 were obtained with GMOS on the 8.1m Gemini South
telescope as a part of fast turnound program GS-2016B-FT-17. The GMOS
images are discussed in Section~\ref{sec:DESJ2346-5203}.

The number of DES targets \citep[selected from SV, Y1,
Y3][]{DESSV,DESY1,DESDR1} observed in each run is given in Table~1.
Non DES targets were also observed and those are described elsewhere
\citep[e.g.,][]{Sch++17,Wil++18,Ost++18,Agn++18a}.  Since both imaging
and spectroscopy are generally required for confirmation, and our runs
were tightly scheduled during the Fall 2016 DES visibility season, we
adopted a running prioritization scheme. Observations of hitherto
unobserved candidates were random, subject to airmass
constraints. Once a candidate was observed, either in imaging or in
spectroscopy, a quick assessment was made, generally the night of the
observations or the next day. If the candidate could be ruled out
based on the available data it was dropped from the target list. If it
was confirmed, or considered promising (e.g. NIQ in spectroscopy or
two point sources with an extended source in the middle in imaging),
its priority was raised for the subsequent complementary runs.  The
coordinates and follow-up outcome of targets observed during the Fall
2016 are given in Tables~2 and~3 for two of the selection techniques
and in companion papers II and III for the other techniques.

\begin{table*}
\caption{Summary of Fall 2016 Campaign Observing Runs}
\label{tab:or}
\begin{tabular}{llllll}
Dates         & Telescope & Instrument        & PI                 &DES Targets Observed & Notes \\
Sep 20-21 & Keck-2     & NIRC2                & Treu             &0 & Technical issues and weather losses\\
Sep 25-28 & NTT          & EFOSC2              & Anguita       &37&\\
Oct 29-31 & Shane 3m & ShARCS              & Rusu            & & Lost to weather\\
Nov 19-20 & Keck-2    & ESI                      & Fassnacht   & 8 & \\
Dec 3-6    & NTT          & EFOSC2              & Anguita       & 40& \\
Dec 3-8    & SOAR        & SAMI                  & Motta/Treu & 60 & Poor weather on
                                                     Dec 8\\
Dec 16      & Keck-2     & NIRC2                & Treu             &&
                                                                     Lost
                                                                     to
                                                                     weather

\end{tabular}
\end{table*}

\section{Follow-up of targets selected with the Outlier Selection Technique}
\label{sec:outliers}

Twenty-six targets selected with the outlier selection technique (OST)
introduced by \citet{Agn17} were observed during the fall 2016
campaign.  The candidates are listed in Table~\ref{tab:summaryOST}
together with a summary of the follow-up data and outcome.

\NcontAA\ candidates were identified as contaminants, \NIncAA\ could
not be securely classified based on the available data and are thus
considered inconclusive. For one system, the spectroscopy and
morphology are possibly consistent with it being a quadruply imaged
quasar in an usual configuration, although confirmation will require
Hubble Space Telescope or Adaptive Optics imaging, given the small
image separation of the system. The system is described in detail in
the next section (\ref{sec:DESJ2346-5203}) along with a potential lens
model. The success rate of this sample ranges between 0 and 42\%,
depending on how many of the inconclusive candidates are actual lens
systems, including the possible quad.

Two classes of contaminants stand out: i) low redshift star forming
galaxies (7/26, i.e. 27\%); QSO+star pair (at least 6/26,
i.e. 23\%). Both classes of objects are expected to be potential
contaminants in photometrically selected samples \citep{Agn++15,WAT17}
and improved algorithms are required to reduce this source of
contamination. QSO+star pairs were also common contaminants in
SQLS. Low redshift star forming galaxies were less common in SQLS
probably by virtue of the spectrosopic preselection and the
availability of $u$-band photometry in SDSS.

Overall, this method did not produce any confirmed lens during
  the Fall 2016 campaign, even though it has been applied with success
  to other datasets \citep{Agn++18a,Agn++18c}. Given the small numbers
  of targets involved, the low yield in this campaign could simply be
  a statistical fluctuation. In any case, there is certainly scope for
  improving the rejection of contaminants noted above.

\begin{table*}
\caption{Summary of observed targets selected with the Outlier Selection Technique}
\label{tab:summaryOST}
\begin{tabular}{lcccl}
\hline
ID                                         & $i$ Mag   & SpecObs  & ImaObs        & Notes \\
\hline
\footnotesize
DESJ234628.18-520331.6 & 20.00 & NTT 9/21, 9/25 & GEMINI 12/6     & Poss. a quad; $z_d=0.48$, $z_s=1.87$ \\
                                            &           &  12/4, 12/5         & SOAR 12/6     & \\

DESJ024326.34-151729.8 & 20.01 & NTT 12/4  &               & Inconc. Two faint traces, no strong emission lines \\
DESJ030539.52-243459.8 & 19.27 &                 & SOAR 12/7       & Inconc. Two point sources with no AO and $0.9''$ seeing\\
DESJ042316.01-375855.4 & 19.89 & NTT 12/3  & SOAR 12/4     & Inconc. Broad emission line at 4967\AA QSO, spectrally unresolved;\\
                                            &           &                  &                        &  SOAR one point source + something extended \\
DESJ042407.95-593806.2 & 19.46 &                  & SOAR 12/4     & Inconc. Point source + something extended \\
DESJ054454.27-471138.1 & 20.61 &                  & SOAR 12/6     & Inconc. SOAR two objects or elongated. bad seeing 0$\farcs$9\\
DESJ061553.23-600552.9 & 18.96 & NTT 9/27  &               & Inconc. QSO z =1.66 unresolved\\
DESJ061838.92-495007.7 & 19.78 &                  & SOAR 12/6     & Inconc. Point source + something extended \\
DESJ065959.89-563521.0 & 19.33 &                  & SOAR 12/3     & Inconc. Point sources or galaxies? \\
DESJ224752.94-431515.4 & 20.33 & NTT 12/5  & SOAR 12/3     & Inconc. Two point objects; QSO at z=0.74 + something faint \\
DESJ220006.63-634447.8 & 19.03 & NTT 9/27  &               & Inconc. QSO $z=1.63$ + faint unidentified trace \\
\hline
DESJ004714.95-204838.5 & 19.21 & NTT 9/25  &               & Contam. $z=0$ star forming galaxy \\
DESJ005426.19-240434.0 & 19.55 & NTT 12/5  & SOAR 12/3    & Contam. Two point sources + galaxy? Emission line galaxy \\
                                            &           &                  &                               &$z=0.354$ + faint no emission\\
DESJ011753.38-044308.0 & 18.60 & NTT 9/26  &               & Contam. Star forming $z=0.138$ \\
DESJ021722.30-551042.2 & 17.29 & NTT 9/26  &               & Contam. QSO $z=1.08$ + star (based on Mg5175 and NaD) \\
DESJ034150.96-572205.8 & 19.70 & NTT 9/26  &               & Contam. QSO $z=1.19$ + featureless spectrum (likely a star) \\
DESJ043949.66-564319.8 & 19.85 & NTT 12/4  &               & Contam. Emission line galaxy at $z=0.351$\\
DESJ045152.71-534504.9 & 18.43 & NTT 9/26  &               & Contam. QSO $z=1.21$ + Star \\
DESJ051207.72-222213.3 & 19.04 & NTT 12/3  & SOAR 12/3     & Contam. Narrow line AGN at $z=0.350$ +featureless trace\\
DESJ200531.34-534939.3 & 19.21 & NTT 9/25 9/26 &           & Contam: Two traces, one QSO at $z=1.73$ \\
                                            &            &                  &               & + featureless (likely a star) \\
DESJ204725.72-612846.7 & 20.15 & NTT 9/27  &               & Contam: QSO z=0.93 (single line at 5379\AA) plus star  \\
DESJ214123.97-592705.8 & 19.60 & NTT 9/25  &               & Contam. Two traces: emission galaxy and absorption line\\
                                           &           &                   &                &  companion at $z=0.47$ \\
DESJ220501.19+003122.9 & 19.62 & NTT 9/26  &               & Contam. QSO $z=1.65$ +faint but different trace \\
DESJ230317.10-454136.8 & 17.82 & NTT 9/27  &               & Contam. Star forming galaxy at $z=0.097$\\
DESJ233411.19-642139.9 & 20.80 & NTT 12/5  & SOAR 12/4     & Contam. SOAR two point sources; Faint unresolved \\
                                           &            &                   &               & [OII] [OIII] emission at $z=0.60$\\
DESJ233520.73-464618.9 & 18.24 & NTT 9/27  &               & Contam. QSO $z=1.65$ + Star\\
\end{tabular}
\end{table*}

\subsection{A candidate quad. Lens models and discussion of the case of DESJ2346-5203} \label{sec:DESJ2346-5203}

\begin{figure*}
  \centering
  \includegraphics[angle=0, width=\textwidth]{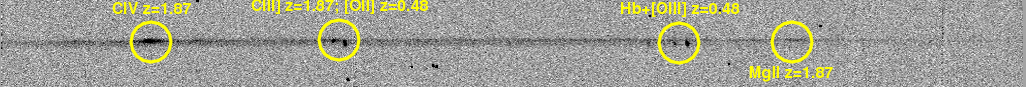}
  \includegraphics[angle=0, width=\textwidth]{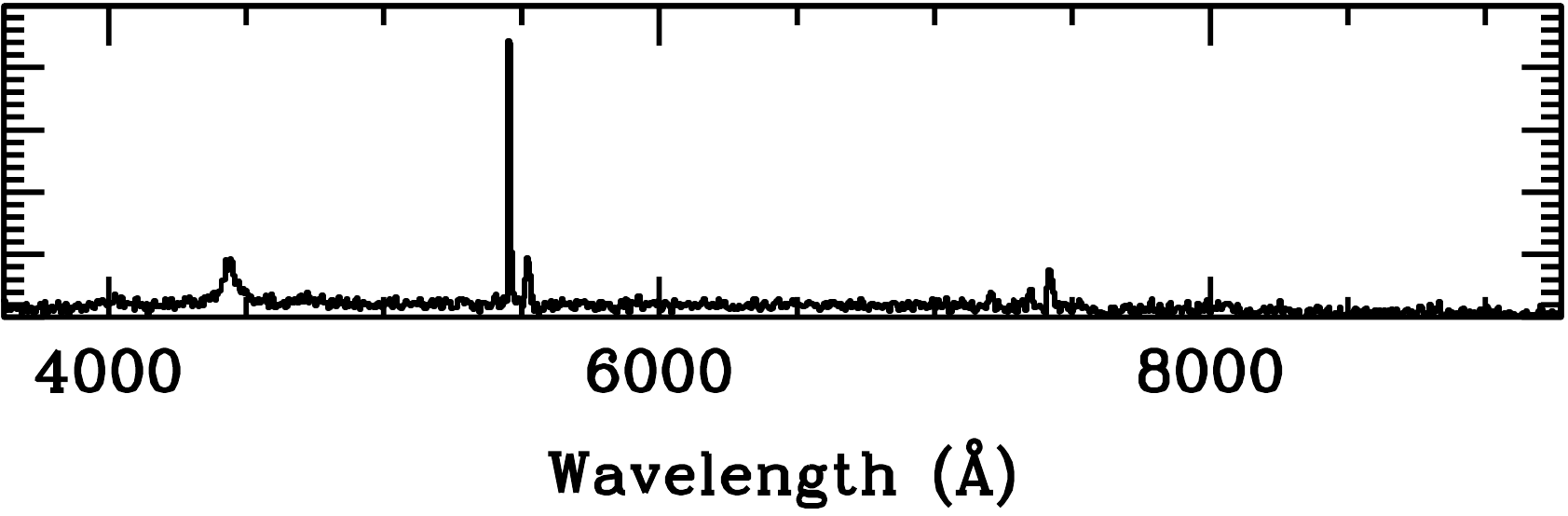}
  \caption{NTT spectrum of DESJ2346-5203 taken on September 25
    2016. The single spectrum clearly shows emission lines at multiple
    redshifts. Top: 2D spectrum. Note the broad CIV and MgII emission and the different spatial
    extent of CIII] and [OII]. Bottom: extracted 1D spectrum.}
\label{fig:NTT_spec_D2346}
\end{figure*}

The NTT spectra of DESJ2346-5203 (Figure~\ref{fig:NTT_spec_D2346}) are
consistent with a small separation (subarcsecond separation) lens. The
candidate deflector is an emission line galaxy at $z_d=0.48$, while
the source is a QSO at $z_s=1.87$. The distance between the two traces
is approximately 2-3 pixels (i.e. $0\farcs48-0\farcs72$) along the
slit (position angle 20$\deg$ East of North). Unfortunately, the
resolution of the DES imaging data was not sufficient to confirm it as
a lens. Therefore, we obtained high resolution imaging data (3x263s
exposures) using the Gemini South Telescope in excellent seeing,
through a fast turnaround program.

In order to investigate whether the system is quadruply imaged or not,
we fit a lens model to the GEMINI GMOS-S data. We use a singular
isothermal ellipsoid with additional external shear as the deflector
mass model, elliptical Sersic profiles for the extended source galaxy
and the lens galaxy and point sources with fixed relative
magnifications based on the lens model for the quasar images. The PSF
is estimated from a bright star in the image. We note that the PSF of
the exposure is highly elliptical. The modeling is performed with the
lens model software \textit{lenstronomy} \citep[based on][available at
\url{https://github.com/sibirrer/lenstronomy}]{BirrerEtal2015,B+A18}.

\begin{figure*}
  \centering
  \includegraphics[angle=0, width=180mm]{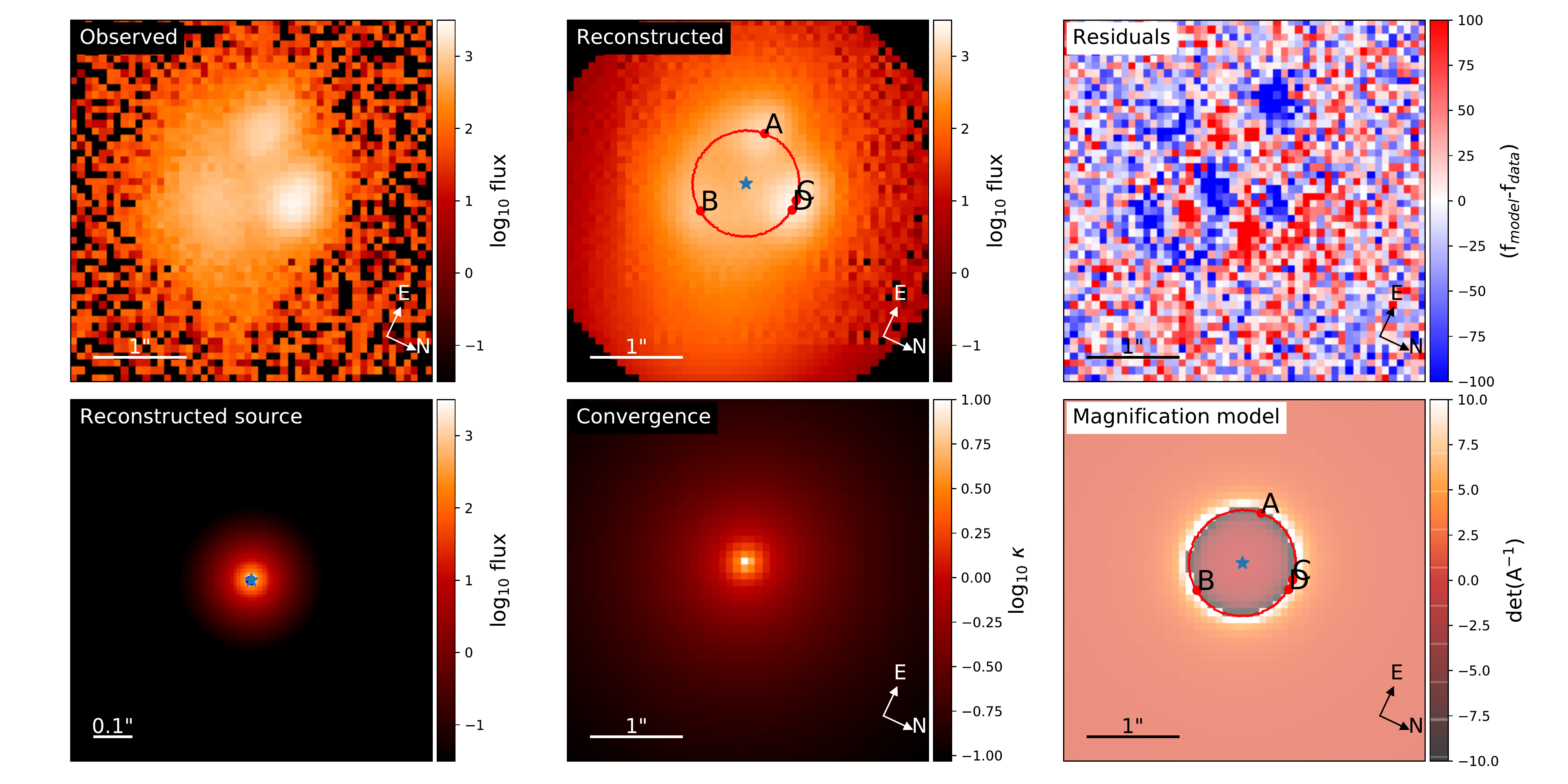}
  \caption{A lens model for the quad candidate DESJ2346-5203. \textbf{Upper left:} The reduced GEMINI image data. \textbf{Upper middle:} Image reconstruction. \textbf{Upper right:} Normalized residuals of the model compared to the data. \textbf{Lower left:} Source reconstruction. \textbf{Lower middle:} Lens light reconstruction. \textbf{Lower right:} Magnification model.}
\label{fig:lens_model_2346}
\end{figure*}

The lens model reproduces the image configuration reasonably
well. Figure~\ref{fig:lens_model_2346} shows a possible lens model of
the candidate DESJ2346-5203. Interestingly, the reconstruction
requires extra flux at the position of the deflector, consistent with
the detection of the deflector.  The lens model is almost spherical to
match the rather unusual image configuration of two very nearby bright
images (C \& D) with a circularized Einstein radius of approximately
$0\farcs61$. The small image separation and the emission line
  suggest that if the system is a lens, the deflector is a late-type
  galaxy, and thus not one of the more common early-type deflectors
  considered in our forecasts.

The model is also consistent with a non-zero extended source that
forms an Einstein ring configuration, although much fainter in
brightness than the point sources. There are residuals left between
the reconstructed model and the image. Some of them can be attributed
to a potentially anomalous flux ratio between the point sources due to
micro-lensing of stars. We also attempted to model the system as a
doubly-imaged quasar (images A and C+D being a single image) +
quadruply image host galaxy, similar to the case of SDSSJ1206+4332
\citep{Agn++16}. No good fit could be found for relatively simple mass
models like elliptical power laws.

Additional information can be gathered by deconvolving the Gemini
images using techniques developed by one of us (F.~C.). The
deconvolved image shows that image A is consistent with being point
like while images CD are well described by a single point source, in
the sense that if they are two merging images they must be unresolved
at the resolution of this image. Image B is not well described by a
point source. Subtracting the point sources in the deconvolved images
does not show a significant excess consistent with a putative lens
galaxy, although of course it could be present below the noise level.

Based on the spectroscopy, the lens model, and the deconvolved images,
we conclude that DESJ2346-5203 is unlikely to be a strong lens system
in a simple traditional configuration (e.g. four images of a quasar
with a galaxy in between). The lens model leaves substantial residuals
and is somewhat contrived with images C and D being practically on top
of each other. However, we note that unusual morphologies
\citep{O+M09}, or cases with extreme flux ratio anomalies like the one
presented by \citet{Lin++17} are difficult to rule out (or identify!)
without higher resolution imaging or spatially resolved spectroscopy.

\begin{table*}
\caption{Summary of observed targets selected with the Morphological algorithm}
\label{tab:summarymorph}
\begin{tabular}{lcccl}
\hline
ID                                          &  Mag    &   SpecObs   &  ImaObs         &  Notes \\
\hline
\footnotesize
DESJ052553.73-555937.1 & 20.16  & SOAR 12/5 & SOAR 12/7 & Inconc. SOAR Two point sources. Single narrow emission line at 6894\AA\ \\
& & & & but spectrum nesolved into 2 components\\
\hline
DESJ003848.42-480147.7 & 20.54  & NTT 12/5 & SOAR 12/4 &  Contam. SOAR: two point sources. Two stars\\
DESJ013037.61-535419.0 & 21.05  & NTT 12/3 & SOAR 12/3 &  Contam. SOAR: two point sources. Two stars\\
DESJ025629.40-413712.6 & 20.59  & NTT 12/3 &  &  Contam. Emission line galaxy\\
DESJ031908.53-410629.4 & 20.30  & NTT 12/5 & SOAR 12/3 &  Contam. SOAR: two point sources. Two stars\\
DESJ032730.55-402712.0 & 20.17  & NTT 12/5  & SOAR 12/7 & Contam. SOAR: two point sources. Single MgII emission line, probably a \\
& & & & quasar pair, $9079$ and $z=0.9021$\\
DESJ040352.63-450052.3 & 19.98  & NTT 12/5 & SOAR 12/3 &  Contam. SOAR: two point sources. QSO z=2.28 +star\\
DESJ040934.96-521619.8 & 19.11  & NTT 12/5 & SOAR 12/4 &  Contam. SOAR: two point sources. Two stars\\
DESJ044538.42-582847.0 & 19.82  & NTT 12/5 & SOAR 12/5 &  Contam. SOAR: two point sources. Two stars\\
DESJ045613.66-582519.6 & 20.10  & NTT 12/4 &  &  Contam. galaxies\\
DESJ050713.60-584440.0 & 20.53  & NTT 12/4 & SOAR 12/4 &  Contam. SOAR: two point sources, Two stars\\
DESJ051340.97-425352.5 & 19.35  & NTT 12/4 & SOAR 12/3 &  Contam. SOAR: two point sources. QSO + star\\
DESJ051813.72-434216.3 & 19.06  & NTT 12/4 & SOAR 12/4 &  Contam. SOAR: two point sources. QSO +star\\
DESJ053232.64-445432.7 & 19.04  & NTT 12/3 &  & Contam. Emission line galaxy + faint object\\
DESJ053239.19-584823.0 & 20.37  & NTT 12/3 &  &  Contam. Two stars\\
DESJ061727.03-482426.9 & 18.91  & NTT 12/5 & SOAR 12/3 &  Contam. Two stars\\

\end{tabular}
\end{table*}

\section{Follow-up of targets selected with the Morphological Algorithm}
\label{sec:morphological}

We also followed up a set of candidates identified via a morphological algorithm that was originally developed by two high-school students \citep{sivakumar15} to search for quasars in the Sloan Digital Sky Survey. This algorithm uses a set of morphological cuts followed by the application of image segmentation algorithms to find lensed quasar candidates. An initial set of objects were selected by applying the following criteria to all objects from the DES Y1A1\_COADD \citep[][Y1 means year one]{morganson18}.
\begin{itemize}
\item Dec $> -60\deg$ to avoid the Magellanic Clouds.
\item In order to eliminate extended sources we require that the Petrosian radius be less than 5 pixels, i.e. $1\farcs35$.
\item To select objects with quasar-like colors we then apply color cuts $-0.2495 < g-r < 0.3393$, $-0.3158 < r-i < 0.658$, $-0.239 < i-z < 0.591$,  similar\footnote{The $r-i$ cut is slightly different than the original one due to a small computing error.} to those implemented by \citet{richards01} and converted to DES colors using the equations in Appendix A.4 of \citet{DESY1}. All magnitudes are $MAG\_AUTO$ as calculated by SExtractor \citep{bertin96,bertin02}.
\item We require $17 < g < 22$ and $17 < r < 22$ . The upper cut eliminates saturated objects and the lower one removes faint galaxies that can be misclassified as stars.
\item The object detection in DES \citep{bertin96,bertin02} does not de-blend the individual components of small image separation lensed quasars into separate objects. These blended objects appear as extended sources and can be identified by requiring that the magnitude measured assuming a stellar profile, $MAG\_PSF$,  be different from $MAG\_AUTO$, namely $ABS(r_{MAG\_AUTO} - r_{MAG\_PSF} )> 0.12$.
\item Finally we require that the objects have $FLAGS\_G =1$ or $FLAGS\_G = 3$. This selects objects that have neighbors or neighbors and blended. This eliminates the many objects that are isolated. Additionally we require $FLAGS\_G < 4$ and $FLAGS\_R < 4$ to eliminate objects that contain any saturated pixels.
\end{itemize}

These cuts select 112,820 candidates. We then obtain postage stamp images of each candidate and run image segmentation algorithms on them to identify individual components in the images. Two algorithms were used for this step, the marker-controlled watershed \citep{beucher92} and the random walker \citep{grady06}, with implementations modified from those in the Python scikit-image package \citep{scikit-image}. The marker-controlled watershed algorithm operates on binary images so the color postage stamps were first converted to black and white using adaptive thresholding. A distance function was defined to identify seeds in the image that correspond to the images to be extracted. These seeds provide the locations from which the algorithm floods the image to find distinct boundaries, and this method avoids over-segmentation of the image. This algorithm is efficient at finding the seed locations but does not provide the most accurate segmentation. So for images that were successfully segmented by the watershed method we then applied the random walker algorithm to them. The random walker requires color images and starts with a seed and then expands outwards to look for neighbors to segment the image. The seeds from the watershed algorithm are used as the starting points for the random walker. The final segmented images and their properties are obtained from the random walker algorithm as it provides accurate segmentation with clear boundaries. After the image segmentation step we are left with 70,823 candidates. These candidates were visually inspected and reduced to 156.

We then applied a second set of color cuts that incorporate the W1 and W2 bands from WISE (Eq.~1; using a matching radius of $1''$), based on Figure~3 of \citet{Ost++17}, to further narrow down the sample. The WISE magnitudes are in the Vega system. The conversions for the WISE data are $W1_{AB} = W1_{Vega} + 2.699$ and $W2_{AB} = W2_{Vega} + 3.339$ which are given by \citet{jarrett11}.  Candidates for which the value of $W1$ is an upper limit were also removed as their colors are not reliable.

\begin{eqnarray}
-0.5 < (i - W1) < 2.5\nonumber\\
-0.2 < (g-i) < 1\\
-0.1 < (W1-W2) < 1.2\nonumber
\label{eq:colorcuts}
\end{eqnarray}

These final cuts yielded 35 candidates which were all then visually
inspected to select the final sample of 18 candidates for
spectroscopic follow-up. We were able to observe 16 of the 18
candidates and these are listed in Table~\ref{tab:summarymorph}
together with a summary of the follow-up data and outcome.  In short,
one of the candidates remains inconclusive, and will require higher
resolution spectroscopy or deeper imaging to finalize its
classification. Fifteen objects are found to be contaminants,
including 8 star pairs, 3 QSO+star pairs, 1 probable QSO pair, and 3
galaxies. Based on the performance so far it is clear that this
  method requires further improvements, especially in the rejection of
  stellar contaminants, in order to be competitive with other methods
  in terms of purity.

\section{Summary Statistics of the 2016 STRIDES Follow-up Campaign}
\label{sec:ss}

In addition to the Outlier Selection Technique introduced by
\citet{Agn17}, and the morphological technique described in
Section~\ref{sec:morphological}, other techniques were developed by
members of the STRIDES collaboration. Their selection techniques and
results of follow-up are described in other papers of this series
\citep[][and Ostrovski et al. 2018, in preparation]{Ang++18}.
Overall, taking into account all selection methods, \NOBSTOT\
DES-selected candidates were observed. \NlensTOT\ were confirmed as
lensed quasars, including \NquadTOT\ quads, \NNIQTOT\ were classified
as NIQs. For \NIncTOT\ the observations were inconclusive, and the
rest were rejected as contaminants.

The scale of the follow-up is sufficient to get a first
  assessment of the success rate of our candidate selection
  techniques, and compare it with previous searches. The overall
success rate across all techniques is in the range 6-35\%.  This is a
good success rate considering that the selection is purely photometric
and no spectroscopic pre-selection is applied.  For comparison, the
most recently completed large scale search for lensed quasars is the
Sloan Digital Sky Survey Quasar Lens Search
\citep[SQLS;][]{Ina++12}. Starting from a sample of 50,836
spectroscopically confirmed quasars, they identified 520 candidates
based on color and morphology. Thirty (including 26 in the so-called
``statistical sample'') of those were confirmed as lensed quasars. One
important class of contaminants were 81/520 QSO pairs,
i.e. approximately 16\%. Another important class of contaminants were
QSO+star (at least 100), to which one should probably add most of the
objects classified as ``no lens'' based on imaging data (158;
spectroscopic classification is not available for this class; these
are most likely to be QSO+star, Oguri 2017, private communication), A
few objects could not be confirmed as lenses due to small separation
(9), although some of them could very well be lenses. Thus, the
overall success rate is at least 6\% but possibly a little
higher. QSO+star class comprises at least 19\% of the spurious
candidates, and perhaps as high as 50\%.  We refer to the individual
papers of this series for a breakdown in the various class of
contaminants for the STRIDES searches.

A more recent search is that carried out by the SDSS-III BOSS quasar
lens survey \citep[BQLS;][]{Mor++16}. Similarly to SQLS, they start from
spectroscopically confirmed quasars and look for evidence for
lensing. In their initial study, they confirmed as lenses 13 of their
55 best candidates, i.e. a success rate of 20\%. Of the top 55
candidates 11 are confirmed quasar pairs, some of which might be
unrecognized lenses.

In addition, we can compare the number of forecasted lenses with
  the number of confirmed lenses to roughly estimate the completeness
  of our search so far, keeping in mind that the searches were
  conducted on partial and different DES data releases. The two
search algorithms presented in this paper were applied to the Y1A1 DES
data release, which covers approximately 1800 square degrees,
i.e. 36\% of the DES footprint, shallower than full depth. The
algorithm presented in paper II \citep{Ang++18} was applied to the
Y1+Y2 footprint, corresponding to approximately the entire DES
footprint, shallower than full depth.  The algorithm presented in
paper III (Ostrovski et al. 2018) was applied to the part of the Y3
data release that overlaps with the VISTA-VHS survey (approximately
half the entire DES footprint, shallower than full depth).

\begin{figure}
  \centering
  \includegraphics[angle=0, width=\columnwidth]{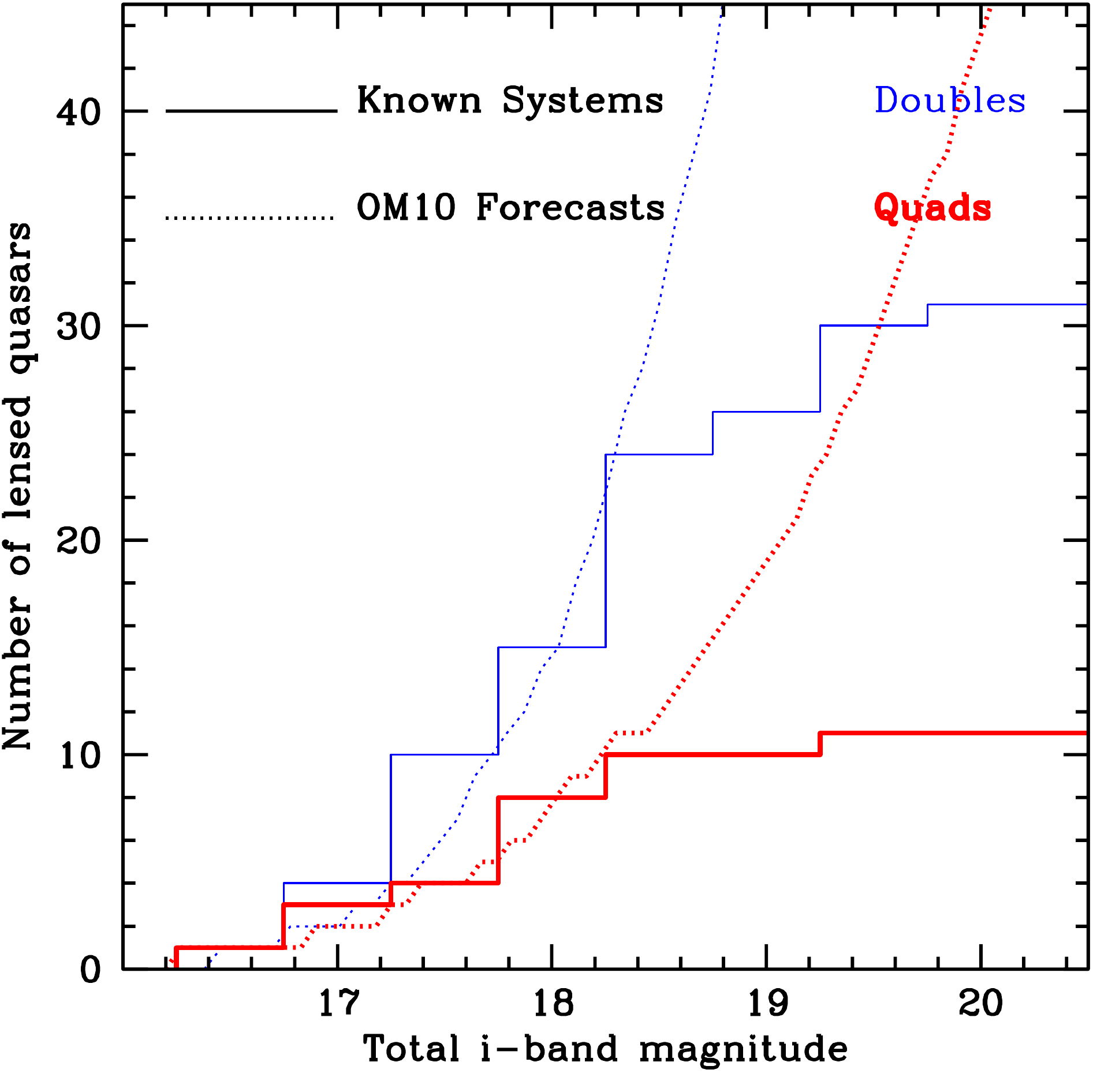}
  \caption{Comparison between known lenses (including those discovered before and after the Fall 2016 STRIDES campaign) within the DES footprint (solid lines) and OM10 forecasts (dotted lines). The thin blue lines indicate doubles (excluding NIQs), and the thick red lines indicate quads. The vertical axis shows the cumulative number of lenses, while the horizontal axis shows the total $i$-band magnitude measured within a $5''$-diameter aperture in DES images.}
\label{fig:complot0}
\end{figure}

Considering only the brighter systems ($i\sim20.2$ or brighter) that
should have been detectable at reduced depth, we expected
(\S~\ref{sec:forecasts} roughly 60 lensed quasars, including 15
quads. We confirmed \NlensTOT\ lenses, including 2 quads (possibly 8/3
if one wishes to include DESJ2346-5203). It is unlikely that more
quads are hiding amongst the 33 inconclusive systems (including NIQs),
as those generally tend to be easily to confirm due to their peculiar
morphology. Thus, we conclude that a large fraction of quads (and
possibly doubles) brighter than $i\sim20.2$ remains to be found in the
DES footprint, motivating additional searches in subsquent DES data
releases and follow-up campaigns. This conclusion is consistent with
the discovery of doubles and quads in the DES footprint, before
\citep{Agn++15,Ost++17,Lin++17} and after \citep{Agn++18b} the
conclusion of the Fall 2016 campaign. At the moment of this
  writing, considering all known lensed quasars within the DES
  footprint including those discovered before and after the STRIDES Fall 2016
  campaign, there is a good agreement between the forecasts and the
  observations for $i\lesssim18.5$ (see
  Figure~\ref{fig:complot0}). Beyond this limit the number of known
  lensed quasars increases much more slowly than forecasted,
  suggesting that many lenses remain to be found.

The public data releases \citep{GaiaDR2} of the Gaia
  Satellite \citep{Gaia16} have provided another powerful tool in the
  arsenal of the lens quasar finding community. Gaia's high resolution
  positions and proper motions have been shown to be extremely
  powerful by themselves \citep{Kro++18} and especially in combination
  with optical and mid-IR images for identifying lensed quasars and
  reject contaminants \citep{Agn++18b,Lem++18,A+S18}. The fast
  turnaround of these discoveries after the data releases is very
  encouraging for STRIDES both in terms of the prospects of
  completeness and success rate of targeted follow-up.

\begin{figure}
  \centering
  \includegraphics[angle=0, width=\columnwidth]{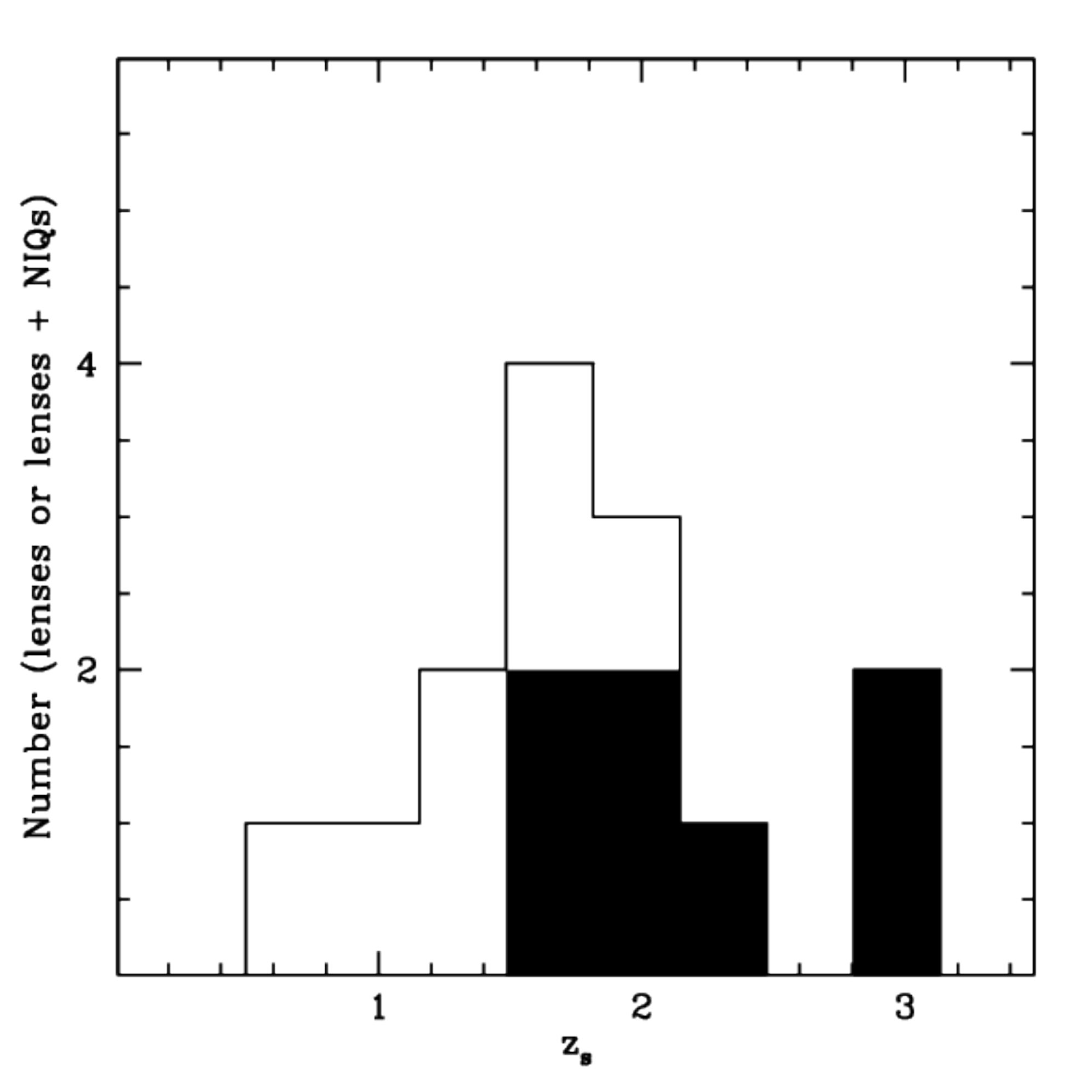}
  \caption{Distribution of quasar redshifts for confirmed lenses (shaded histogram) and NIQs (open histogram).}
\label{fig:plottazs}
\end{figure}

Finally, we can make a further comparison between the forecast
  and the properties of entire sample, by looking at the quasar
  redshift distribution. Given the small number statistics we combine
  both confirmed lenses and NIQs, assuming that they are drawn from
  the same distribution, even though this of course will need to be
  revisited at the end of the STRIDES multi-year effort. The
  distribution is shown in Figure~\ref{fig:plottazs}. As forecasted,
  the distribution peaks at $z_s\sim2$, and drops off below 1 and
  above 3. Whereas the numbers are still too small for a quantitative
  comparison between forecast and detections, the qualitative
  agreement is encouraging, especially because contrary to the SDSS
  searches we did not rely on u-band imaging or spectroscopic
  information for selection of candidates.

 \section{Summary}
\label{sec:summary}

We have presented an overview of the STRIDES program, an external
collaboration of the Dark Energy Survey aimed at finding and studying
strongly lensed quasars, and outlined some of the results of the first
comprehensive follow-up campaign.  The main results of this paper can
be summarized as follows:

\begin{itemize}
\item Our detailed forecasts indicate that about 50 quads and 200 doubles should be detectable in DES data.  Of those, approximately 60 should be bright enough for time delay measurements with 1-2m class telescopes, while the rest will require a 4m class telescope for monitoring. All the systems will be bright enough to measure stellar velocity dispersion with 8-10m class telescopes.
\item The STRIDES lens classification scheme is presented. In addition to confirmed lenses, and inconclusive systems, we adopt the class of Nearly Identical Quasars (NIQ) to identify inconclusive targets which are particularly promising for additional follow-up.
\item We detail the results of the follow-up of 42 targets selected by two of the search techniques (Outlier Selection and Morphological; OST and MT respectively).  One of those is a candidate quadruply imaged quasar (DESJ2346-5203; see the next bullet item), 11 are inconclusive, and and 30 are contaminants. The contaminants are dominated by QSO+star pairs for the OST and by star pairs for MT.
\item Based on the analysis of $0\farcs4$ seeing Gemini-S images of the candidate quad DESJ2346-5203, and we conclude that this is not a quadruply imaged quasar in a classic configuration. If it is a multiply imaged quasar the morphology requires a complex deflector or extreme flux ratio anomalies. High resolution imaging or spectroscopy are required to definitely rule out (or confirm) this system as a lens.
\item We summarize the results of our Fall 2016 observing campaign with the Keck, SOAR, NTT, Shane Telescopes. In total we followed up \NOBSTOT\ targets, confirming \NlensTOT\ lenses including \NquadTOT\ quads, and found \NNIQTOT\ NIQs. The observations were inconclusive for \NIncTOT\ targets, yielding a success rate in the range 6-35\%. This success rate is comparable with those of other large searches for lensed quasars even though no spectroscopic information nor $u$-band imaging was available to help in the selection.
\end{itemize}

In conclusion, the results of the first extensive STRIDES follow-up
campaign demonstrate that multiply imaged quasars can be found
efficiently from wide field imaging survey even in the absence of
$u$-band or spectroscopic preselection. At the conclusion of our
multi-year campaign, with the investment of telescope time to carry
out imaging and spectroscopic follow-up of DES-selected targets,
STRIDES should more then double the current a sample of known
lensed quasars, and thus enable significant progress in our
understanding of the nature of dark matter and dark energy.

\section*{Acknowledgments}

The authors thank the referee for a constructive report that
  helped improved the paper.  T.T. thanks Masamune Oguri for several
useful conversations on finding lensed quasars and on SQLS. T.T. and
V.M. acknowledge support by the Packard Foundation through a Packard
Research Fellowship to T.T. T.T.  acknowledges support by the National
Science Foundation through grants AST-1450141 and
AST-1714953. C.D.F. and G.C.F.C. acknowledge support from the National
Science Foundation through grants AST-1312329 and
AST-1715611. T. A. and Y.A. acknowledge support by proyecto FONDECYT
11130630 and by the Ministry for the Economy, Development, and
Tourism’s Programa Inicativa Cient\'{i}fica Milenio through grant IC
12009, awarded to The Millennium Institute of Astrophysics
(MAS). F.~C., V.~B., and J.~C. acknowledge support from the Swiss
National Science Foundation. S.H.S. thanks the Max Planck Society for
support through the Max Planck Research Group.

Funding for the DES Projects has been provided by the U.S. Department of Energy, the U.S. National Science Foundation, the Ministry of Science and Education of Spain,
the Science and Technology Facilities Council of the United Kingdom, the Higher Education Funding Council for England, the National Center for Supercomputing
Applications at the University of Illinois at Urbana-Champaign, the Kavli Institute of Cosmological Physics at the University of Chicago,
the Center for Cosmology and Astro-Particle Physics at the Ohio State University,
the Mitchell Institute for Fundamental Physics and Astronomy at Texas A\&M University, Financiadora de Estudos e Projetos,
Funda{\c c}{\~a}o Carlos Chagas Filho de Amparo {\`a} Pesquisa do Estado do Rio de Janeiro, Conselho Nacional de Desenvolvimento Cient{\'i}fico e Tecnol{\'o}gico and
the Minist{\'e}rio da Ci{\^e}ncia, Tecnologia e Inova{\c c}{\~a}o, the Deutsche Forschungsgemeinschaft and the Collaborating Institutions in the Dark Energy Survey.

The Collaborating Institutions are Argonne National Laboratory, the University of California at Santa Cruz, the University of Cambridge, Centro de Investigaciones Energ{\'e}ticas,
Medioambientales y Tecnol{\'o}gicas-Madrid, the University of Chicago, University College London, the DES-Brazil Consortium, the University of Edinburgh,
the Eidgen{\"o}ssische Technische Hochschule (ETH) Z{\"u}rich,
Fermi National Accelerator Laboratory, the University of Illinois at Urbana-Champaign, the Institut de Ci{\`e}ncies de l'Espai (IEEC/CSIC),
the Institut de F{\'i}sica d'Altes Energies, Lawrence Berkeley National Laboratory, the Ludwig-Maximilians Universit{\"a}t M{\"u}nchen and the associated Excellence Cluster Universe,
the University of Michigan, the National Optical Astronomy Observatory, the University of Nottingham, The Ohio State University, the University of Pennsylvania, the University of Portsmouth,
SLAC National Accelerator Laboratory, Stanford University, the University of Sussex, Texas A\&M University, and the OzDES Membership Consortium.

Based in part on observations at Cerro Tololo Inter-American Observatory, National Optical Astronomy Observatory, which is operated by the Association of
Universities for Research in Astronomy (AURA) under a cooperative agreement with the National Science Foundation.

The DES data management system is supported by the National Science Foundation under Grant Numbers AST-1138766 and AST-1536171.
The DES participants from Spanish institutions are partially supported by MINECO under grants AYA2015-71825, ESP2015-88861, FPA2015-68048, SEV-2012-0234, SEV-2016-0597, and MDM-2015-0509,
some of which include ERDF funds from the European Union. IFAE is partially funded by the CERCA program of the Generalitat de Catalunya.
Research leading to these results has received funding from the European Research
Council under the European Union's Seventh Framework Program (FP7/2007-2013) including ERC grant agreements 240672, 291329, and 306478.
We  acknowledge support from the Australian Research Council Centre of Excellence for All-sky Astrophysics (CAASTRO), through project number CE110001020.

This manuscript has been authored by Fermi Research Alliance, LLC under Contract No. DE-AC02-07CH11359 with the U.S. Department of Energy, Office of Science, Office of High Energy Physics. The United States Government retains and the publisher, by accepting the article for publication, acknowledges that the United States Government retains a non-exclusive, paid-up, irrevocable, world-wide license to publish or reproduce the published form of this manuscript, or allow others to do so, for United States Government purposes.

(Some of) The data presented herein were obtained at the W. M. Keck
Observatory, which is operated as a scientific partnership among the
California Institute of Technology, the University of California and
the National Aeronautics and Space Administration. The Observatory was
made possible by the generous financial support of the W. M. Keck
Foundation. The authors wish to recognize and acknowledge the very
significant cultural role and reverence that the summit of Maunakea
has always had within the indigenous Hawaiian community.  We are most
fortunate to have the opportunity to conduct observations from this
mountain.  Based in part on observations obtained at the Southern
Astrophysical Research (SOAR) telescope, which is a joint project of
the Minist\'{e}rio da Ci\^{e}ncia, Tecnologia, Inova\c{c}\~{a}os e
Comunica\c{c}\~{a}oes (MCTIC) do Brasil, the U.S. National Optical
Astronomy Observatory (NOAO), the University of North Carolina at
Chapel Hill (UNC), and Michigan State University (MSU). Based in part
on observations obtained at the Gemini Observatory, which is operated
by the Association of Universities for Research in Astronomy, Inc.,
under a cooperative agreement with the NSF on behalf of the Gemini
partnership: the National Science Foundation (United States), the
National Research Council (Canada), CONICYT (Chile), Ministerio de
Ciencia, Tecnolog\'{i}a e Innovaci\'{o}n Productiva (Argentina), and
Minist\'{e}rio da Ci\^{e}ncia, Tecnologia e Inova\c{c}\~{a}o (Brazil).
Based in part on observations made with ESO Telescopes at the La Silla
Paranal Observatory. This publication makes use of data products from
the Wide-field Infrared Survey Explorer, which is a joint project of
the University of California, Los Angeles, and the Jet Propulsion
Laboratory/California Institute of Technology, funded by the National
Aeronautics and Space Administration.

\bibliographystyle{mnras}


\section*{Affiliations}
  $^1$\ucla\\
  $^2$\eso\\
  $^3$\kipac\\
  $^4$Fermi National Accelerator Laboratory, P. O. Box 500, Batavia, IL 60510, USA \\
 $^5$\epfl\\
 $^6$\mit\\
  $^7$Illinois Mathematics and Science Academy, 1500 Sullivan Road, Aurora IL 60506-1067, USA\\
  $^8$University of California-Berkeley, Berkeley, CA 94720\\
$^9$\abello\\
$^{10}$\ioa\\
$^{11}$\asiaa\\
$^{12}$\ucd\\
$^{13}$\ports\\
$^{14}$\valpo\\
$^{15}$\ufrgs\\
$^{16}$\mpa\\
$^{17}$\milchile\\
$^{18}$\tum\\
$^{19}$\subaru\\
$^{20}$ Cerro Tololo Inter-American Observatory, National Optical Astronomy Observatory, Casilla 603, La Serena, Chile\\
$^{21}$ Department of Physics \& Astronomy, University College London, Gower Street, London, WC1E 6BT, UK\\
$^{22}$ Department of Physics and Electronics, Rhodes University, PO Box 94, Grahamstown, 6140, South Africa\\
$^{23}$ Kavli Institute for Cosmology, University of Cambridge, Madingley Road, Cambridge CB3 0HA, UK\\
$^{24}$ Laborat\'orio Interinstitucional de e-Astronomia - LIneA, Rua Gal. Jos\'e Cristino 77, Rio de Janeiro, RJ - 20921-400, Brazil\\
$^{25}$ Observat\'orio Nacional, Rua Gal. Jos\'e Cristino 77, Rio de Janeiro, RJ - 20921-400, Brazil\\
$^{26}$ Department of Astronomy, University of Illinois at Urbana-Champaign, 1002 W. Green Street, Urbana, IL 61801, USA\\
$^{27}$ National Center for Supercomputing Applications, 1205 West Clark St., Urbana, IL 61801, USA\\
$^{28}$ Institut de F\'{\i}sica d'Altes Energies (IFAE), The Barcelona Institute of Science and Technology, Campus UAB, 08193 Bellaterra (Barcelona) Spain\\
$^{29}$ Institut d'Estudis Espacials de Catalunya (IEEC), 08193 Barcelona, Spain\\
$^{30}$ Institute of Space Sciences (ICE, CSIC),  Campus UAB, Carrer de Can Magrans, s/n,  08193 Barcelona, Spain\\
$^{31}$ Department of Physics and Astronomy, University of Pennsylvania, Philadelphia, PA 19104, USA\\
$^{32}$ Centro de Investigaciones Energ\'eticas, Medioambientales y Tecnol\'ogicas (CIEMAT), Madrid, Spain\\
$^{33}$ Department of Astronomy/Steward Observatory, 933 North Cherry Avenue, Tucson, AZ 85721-0065, USA\\
$^{34}$ Jet Propulsion Laboratory, California Institute of Technology, 4800 Oak Grove Dr., Pasadena, CA 91109, USA\\
$^{35}$ Instituto de Fisica Teorica UAM/CSIC, Universidad Autonoma de Madrid, 28049 Madrid, Spain\\
$^{36}$ Department of Astronomy, University of California, Berkeley,  501 Campbell Hall, Berkeley, CA 94720, USA\\
$^{37}$ Lawrence Berkeley National Laboratory, 1 Cyclotron Road, Berkeley, CA 94720, USA\\
$^{38}$ SLAC National Accelerator Laboratory, Menlo Park, CA 94025, USA\\
$^{39}$ Department of Physics, ETH Zurich, Wolfgang-Pauli-Strasse 16, CH-8093 Zurich, Switzerland\\
$^{40}$ Santa Cruz Institute for Particle Physics, Santa Cruz, CA 95064, USA\\
$^{41}$ Center for Cosmology and Astro-Particle Physics, The Ohio State University, Columbus, OH 43210, USA\\
$^{42}$ Department of Physics, The Ohio State University, Columbus, OH 43210, USA\\
$^{43}$ Harvard-Smithsonian Center for Astrophysics, Cambridge, MA 02138, USA\\
$^{44}$ Australian Astronomical Observatory, North Ryde, NSW 2113, Australia\\
$^{45}$ Departamento de F\'isica Matem\'atica, Instituto de F\'isica, Universidade de S\~ao Paulo, CP 66318, S\~ao Paulo, SP, 05314-970, Brazil\\
$^{46}$ Department of Astronomy, The Ohio State University, Columbus, OH 43210, USA\\
$^{47}$ Instituci\'o Catalana de Recerca i Estudis Avan\c{c}ats, E-08010 Barcelona, Spain\\
$^{48}$ Department of Physics and Astronomy, Pevensey Building, University of Sussex, Brighton, BN1 9QH, UK\\
$^{49}$ Department of Physics, University of Michigan, Ann Arbor, MI 48109, USA\\
$^{50}$ School of Physics and Astronomy, University of Southampton,  Southampton, SO17 1BJ, UK\\
$^{51}$ Brandeis University, Physics Department, 415 South Street, Waltham MA 02453\\
$^{52}$ Instituto de F\'isica Gleb Wataghin, Universidade Estadual de Campinas, 13083-859, Campinas, SP, Brazil\\
$^{53}$ Computer Science and Mathematics Division, Oak Ridge National Laboratory, Oak Ridge, TN 37831

\label{lastpage}

\end{document}